\documentclass[
superscriptaddress,
reprint,
showpacs,preprintnumbers,
 amsmath,amssymb,
 aps,
floatfix,
]{revtex4-1}
\usepackage{appendix}
\usepackage{SIunits}
\usepackage{sistyle}
\usepackage[utf8x]{inputenc}
\usepackage{graphicx}   
\graphicspath{{fig/}}   
\usepackage{dcolumn}
\usepackage{bm}
\usepackage{hyperref}
\usepackage{epstopdf}
\usepackage{lipsum}
\usepackage{braket}
\DeclareGraphicsExtensions{.pdf,.png,.jpg,.jpeg,.mps}
\usepackage{pgf}
\usepackage{tikz}

\usepackage{changes}

\newcommand{\sous}[1]{\ensuremath{_{\textrm{#1}}}}

\usepackage[colorinlistoftodos,prependcaption,textsize=tiny]{todonotes}
\begin{document}


\title{Complex temperature dependence of coupling and dissipation of cavity-magnon polaritons from milliKelvin to room temperature}

\author{Isabella Boventer}
\affiliation{Institute of Physics, Johannes Gutenberg University Mainz, 55099 Mainz, Germany}
\affiliation{Institute of Physics, Karlsruhe Institute of Technology, 76131 Karlsruhe, Germany}
\author{Marco Pfirrmann}
\affiliation{Institute of Physics, Karlsruhe Institute of Technology, 76131 Karlsruhe, Germany}
\author{Julius Krause}
\affiliation{Institute of Physics, Karlsruhe Institute of Technology, 76131 Karlsruhe, Germany}
\author{Yannick Sch\"on}
\affiliation{Institute of Physics, Karlsruhe Institute of Technology, 76131 Karlsruhe, Germany}
\author{Mathias Kl\"aui}
\email{Klaeui@Uni-Mainz.de}
\affiliation{Institute of Physics, Johannes Gutenberg University Mainz, 55099 Mainz, Germany}
\affiliation{Materials Science in Mainz, University Mainz, 55099 Mainz, Germany}
\author{Martin Weides}
\affiliation{Institute of Physics, Johannes Gutenberg University Mainz, 55099 Mainz, Germany}
\affiliation{Institute of Physics, Karlsruhe Institute of Technology, 76131 Karlsruhe, Germany}
\affiliation{Materials Science in Mainz, University Mainz, 55099 Mainz, Germany}


\date{\today}

\begin{abstract}
Hybridized magnonic-photonic systems are key components for future information processing technologies such as storage, manipulation or conversion of data both in the classical (mostly at room temperature) and quantum (cryogenic) regime. In this work, we investigate a YIG sphere coupled strongly to a microwave cavity over the full temperature range from $290\,\mathrm{K}$ down to $30\,\mathrm{mK}$. The cavity-magnon polaritons are studied from the classical to the quantum regime where the thermal energy is less than one resonant microwave quanta, i.e. at temperatures below $1\,\mathrm{K}$.  We compare the temperature dependence of the coupling strength $g_{\rm{eff}}(T)$, describing the strength of coherent energy exchange between spin ensemble and cavity photon, to the temperature behavior of the saturation magnetization evolution $M\sous{s}(T)$ and find strong deviations at low temperatures. 
The temperature dependence of magnonic disspation is governed at intermediate temperatures by rare earth impurity scattering leading to a strong peak at $43\,$K. The linewidth $\kappa\sous{m}$ decreases to $1.2\,$MHz at $30\,$mK, making this system  suitable as a building block for quantum electrodynamics experiments. We achieve an electromagnonic cooperativity in excess of $20$ over the entire temperature range, with values beyond $100$ in the milliKelvin regime as well as at room temperature. With our measurements, spectroscopy on strongly coupled magnon-photon systems is demonstrated as versatile tool for spin material studies over large temperature ranges. Key parameters are provided in a single measurement, thus simplifying investigations significantly. 
\end{abstract}
\maketitle
Cavity-magnon polaritons are bosonic quasiparticles associated with the hybridization of a  photon and a magnon - the quanta of a collective spin excitation - within a cavity resonator   \citep{Mills1974}. This hybridization is harnessed for new technologies in data manipulation and processing, and exploits the individual advantages of each component. In the strong coupling regime \citep{Novotny2010,Braumueller2016}, this results in an anticrossing in the dispersion spectrum. The size of the splitting depends on the coupling strength $g_{\rm{eff}}(T)$ as a measure for the frequency of coherent information exchange. The  controllability of quantum coherent systems has already been demonstrated with various resonant platforms \cite{Blatt2008,Hanson2008,Georgescu2014}.  
Hybridized spin ensembles and macroscopic systems such as superconducting qubits are promising approaches for non-linear devices. Long coherence times of spin ensembles are combined with the tunability, scalability and strong coupling to external fields of superconducting quantum circuits  \cite{Xiang2013,Kubo_2011,Imamoglu2007}.
In order to prevent rapid decoherence to an open enviroment, both components are individually strongly coupled to a cavity resonator's photon mode  \cite{Xiang2013}. This resonator mediates the coupling by acting as a quantum bus. It allows for control of coupling between a spin ensemble and qubit, i.e. the memory and data processor, respectively \cite{Xiang2013,Kubo_2011}.
Recently, several experiments with magnons coupled to photons in  microwave resonators in the strong coupling regime were conducted \cite{Zhang_2014,Tabuchi_2014,Bai2015,Cao2015}. Additionally, strong coupling between a qubit and a magnon in a 3d microwave resonator has already been demonstrated \cite{Tabuchi2015}.
However, these experiments were either performed exclusively at room or cryogenic temperature regimes only.

For a full picture of the behaviour of all individual components (magnon, cavity) and in particular the key parameters coupling strength and local dissipation of the hybrid system one needs to close the gap between the classical and quantum regimes. The hybridized cavity photon - magnon states are characterized by the total coupling strength as a measure for the interaction $g_{\rm{eff}}(T)$, which in order to reach the interesting strong coupling regime has to greatly exceed both the magnonic and the photon dissipation factors $\kappa_{\rm{m}}$ and $\kappa_{\rm{l}}$ ($g_{\rm{eff}}(T)\gg\kappa_{\rm{m}}, \kappa_{\rm{l}} $). Thus, for the characterization of the temperature dependence of the magnon polariton properties the study of the coupling strength and the system's dissipation factors is necessary. 
Generally, for a $N$ particle system the coupling strength is proportional to the square root of $N$ ($g_{\rm{eff}}(T)\propto \sqrt{N}$). Hence, it should scale accordingly to the change in $N$ as the temperature is changed \cite{Tavis_1968, Novotny2010, Tabuchi2015}. However, it is unclear if this classical behaviour holds across the whole temperature range. As the second governing factor that determines the coupling regime, we need to ascertain the dissipation processes. For the dissipation, different mechanisms can play a role and in particular opposite temperature dependencies have been observed in the high temperature \cite{Seiden_1964} and the low temperature \cite{Tabuchi2015} regime. This calls for a characterization across the whole temperature range.
Here, we present a temperature dependent study of magnon polaritons from $30\,\mathrm{mK}$  to $290\, \mathrm{K}$.  The microwave photons confined in the standing wave mode of a 3d cavity resonator are strongly coupled to magnons in a sphere made from Yttrium-Iron-Garnet (YIG). The cavity-magnon polaritons are studied across classical to quantum transitions when the thermal energy is less than a single microwave quanta at the resonant transition frequency, i.e. temperatures below $1\,\mathrm{K}$. Specifically, we determine the temperature dependence of the coupling strength $g_{\rm{eff}}(T)$ and the magnon dissipation linewidth $\kappa\sous{m}\left(T\right)$, along with resonance fields, frequencies and cooperativity. Finally, we demonstrate that our measurement scheme can be used to drastically simplify the analysis of magnetic systems by allowing us to efficiently determine a number of key parameters in a single measurement.

In order to understand the properties of magnon polaritons from a theoretical perspective, we start with a
semiclassical description of spin waves. Magnons as bosonic quasiparticles describe collective exictations of spins in materials with a finite magnetic moment \cite{Kittel_1948,ZareRameshti2015}. This is reflected in the precessional motion of the magnetization vector $\textbf{\emph{M}}$. Accordingly, the time evolution is described by the Landau-Lifshitz-Gilbert equation \cite{safo}:
$\frac{\partial \textbf{\emph{M}}}{\partial t}=-\gamma \textbf{\emph{M}} \times \textbf{\emph{H}}_{\rm{eff}} -\frac{\alpha\gamma}{M_{\rm{s}}}\textbf{\emph{M}}(\textbf{\emph{M}}\times \textbf{H}_{\mathrm{eff}}),
$
where $\textbf{\emph{H}}_{\rm{eff}}=\textbf{\emph{H}}_{\rm{ext}}+\textbf{\emph{H}}_{\rm{demag}}+\textbf{\emph{H}}_{\rm{ex}}$ is the effective magnetic field acting on the spins of the sample. 
 $\textbf{\emph{H}}_{\rm{ext}}$ describes the externally applied static field, $\textbf{\emph{H}}_{\mathrm{demag}}$ the internal demagnetizing field and $\textbf{\emph{H}}_{\rm{ex}}$ the exchange field.
The first term characterizes the precession of the electronic spins with $\gamma$ being the gyromagnetic ratio. The second term refers to the intrinsic relaxation, where $\alpha$ and $M_{\mathrm{s}}$ represent the Gilbert damping constant and saturation magnetization, respectively. The sample is brought to saturation by a correspondingly strong, uniform external field. 
A radio frequency magnetic field perpendicular to the external magnetic fields acts as a small perturbation to the spins. 
The lowest order of excitation is the mode of uniformal precession around the axis of saturation magnetization, the so-called Kittel mode \cite{Kittel_1948}. With wave vector $\mathbf{k}=0$ this mode is a special instance of a magnetostatic mode \cite{Fletcher_1959}. The  relation between an external, static field in one direction $\textbf{\emph{H}}=(0,0,H_{\rm{z}})$ and frequency of resonant absorption $\omega_{\mathrm{m}}$, with demagnetizing factors $\textbf{\emph{N}}_{i},i \in({x,y,z})$ is given by \cite{Kittel_1948}:
\begin{equation}
\omega_{\mathrm{m}}=\gamma \sqrt{\left[H_z+\left(N_y-N_z\right)M_z\right]\cdot \left[H_z+\left(N_x-N_z\right)M_z\right]}. 
\end{equation}
For a sphere $N_x=N_y=N_z= \frac{4\pi}{3}$
and the dispersion relation reads:
$
\omega_{\mathrm{m}}=\gamma H_z
\label{H}
$. 
Thus, by sweeping the external field, the resonance frequency $\omega_m$ of the Kittel mode can be tuned until it matches the cavity photon frequency. 
Both the magnonic and the photonic system can be expressed in terms of the harmonic oscillator language \cite{safo} and we use the Tavis-Cummings model for the description of the interaction \cite{Tavis_1968}. It describes  the interaction between a light field, i.e photons and a system of $N$ spins. 
Focusing on the magnon-photon system, we write:
\begin{equation} 
\mathcal{H}_{\rm{TC}}=\hbar\omega\sous{r} a^{\dag}a+\hbar \omega\sous{m}m^{\dag}m+\hbar g_0 \sqrt{2Ns}\left(m^{\dag}a+a^{\dag}m\right),
\label{TC}
\end{equation}
where we set $g_{\rm{eff}}=g\sous{0}\sqrt{2Ns}$ for the experimentally measured total coupling strength, with $s=\frac{5}{2}$ the spin number of the $\mathrm{Fe}^{3+}$ ions in YIG. The first term in Eq. \ref{TC} refers to the photons, the second to the magnons and the third the interaction between both. Each spin couples with a single spin coupling strength
\begin{equation}
g_0=\eta\frac{\gamma}{2}\sqrt{\frac{\mu_{\rm{0}} \hbar\omega_{\rm{r}}}{2\cdot V_{\rm{mode}}}}\,, \label{Eq:coupling_strength}
\end{equation}
where $\eta$ is a factor describing the mode overlap of the photons and magnons within the cavity, $\mu_0$ the vacuum permeability, $\omega_{\rm{r}}$ the resonant frequency of the chosen cavity resonator mode, $V_{\rm{mode}}$ the corresponding mode volume and the factor of $2$ refers to zero point energy of such a cavity \cite{vac}. In case of perfect mode overlap, we could set $\eta=1$.\\
In order to describe the anticrossing within our measured microwave reflection or transmission spectra we utilize the Input-Output formalism \citep{9783540285731}.  It results in complex scattering ratios between the amount of reflected or transmitted amount of energy (output) with respect to the incoming energy (input). 
For the description of our system we consider a single mode excited in the cavity resonator, and that the magnons solely interact with the photons in that specific cavity mode and any direct interaction with the external bath is negligibly small. 
The external bath is formed of discrete modes in the microwave transmission line (feedline). 
Losses of the photon field inside the cavity resonator $\kappa_{\mathrm{i}}$ and externally due to the coupling to one feedline $\kappa_{\rm{e}}$  lead to a total cavity resonator loss $\kappa_{\rm{l}}=\kappa_{\rm{e}}+\kappa_{\rm{i}}$. With these considerations, following the derivation of Ref. \cite{9783540285731}, we obtain for the reflection parameter $\mathcal{S}_{11}$
\begin{equation}
\mathcal{S}_{11}(\omega)=-1 + \frac{2\kappa_{\mathrm{e}}}{\mathrm{i}\left(\omega_{\mathrm{r}}-\omega\right)+\kappa_{\mathrm{l}}+\frac{g_{\rm{eff}}^2}{\mathrm{i}\left(\omega_{\mathrm{m}}-\omega\right)+\kappa_{\mathrm{m}}}},
\label{S11}
\end{equation}
where $\kappa_{\rm{m}}$ refers to relaxation of the magnons. The quantity $\omega_{\rm{r}}$ 
is the resonant frequency of the specific cavity resonator mode and the precession frequency of the magnons in the Kittel mode.
Similarily, we obtain \cite{Tabuchi_2014}
\begin{equation}
\mathcal{S}_{21}(\omega)=\frac{2\sqrt{\kappa_{\mathrm{e,1}}\kappa_{\mathrm{e,2}}}}{\mathrm{i}\left(\omega-\omega_r\right)-\frac{\kappa_{\mathrm{e,1}}+\kappa_{\mathrm{e,2}}+\kappa_{\mathrm{i}}}{2}+\frac{g_{\rm{eff}}^2}{\mathrm{i}\left(\omega-\omega_{\mathrm{m}}\right)-\frac{\kappa_{\mathrm{m}}}{2}}}, \label{S21}
\end{equation}
where the number subscripts in the external loss factors refer to the losses at the two ports, the input and the output coupling of the cavity to the feedline. \\
From the loss parameters, the internal and the external quality factor $Q_{\rm{i}}$
and $Q_{\rm{e}}$,  respectively, can be calculated via $Q=\omega_{\rm{r}}/2\kappa$. 
The quality factor describes the ratio between the stored and per cycle dissipated amount of energy \cite{pozar}.
In order to calculate the average number of photons  $\braket{n}$ within the cavity resonator, the internal quality factor is defined as:
$Q_{\rm{i}}=\omega_{\rm{r}}\frac{\braket{W_r}}{P_{\rm{loss}}},$ where $\braket{W_{\rm{r}}}=\braket{n}\hbar\omega_{\rm{r}}$ denotes the average energy stored in the cavity resonator and $P_{\rm {loss}}$ the power lost due to intracavity resonator losses. 
In combination with expressing the amplitude of the reflected scattering parameter 
$|\mathcal{S}_{\rm{11}}|^2$ as the ratio between backreflected power 
$P_{\rm {o}}$ and input power $P_{\rm{i}}$  and 
$ P_{\rm{loss}}=P_{\rm{i}}-P_{\rm{o}}$, the average photon number is calculated via:
$\braket{n}=\frac{4 P_{\rm{i}}}{\hbar\omega_{\rm{r}}^2}\cdot \frac{Q_{\rm{l}}^2}{Q_{\rm{e}}}$.  
Experimentally, the strong coupling, i.e the  cavity-magnon polariton, is observed in a level repulsion between both coupled harmonic oscillators yielding an avoided crossing (anticrossing)  of the size $2\cdot g_{\rm{eff}}(T)$ in the spectrum \cite{Harder_2016}. \par
For our experiment, we chose a sphere made out of Yttrium-Iron-Garnet ($\rm{Y_{3}Fe_{2}O_{12}}$,YIG) with a diameter of $d=\rm{\SI{0.5}{mm}}$ \cite{ferrisphere}. The sphere has been polished with diamond to a surface quality of $\rm{0.05 \,\mathrm{\mu m}}$. YIG is an insulating ferrimagnet with two antiferromagnetically ordered ferromagnetic sublattices with eight ions at the octahedral and twelve $\rm{\mathrm{Fe}^{3+}}$ ions at the tetrahedral sites with respect to the oxygen lattice sites. This results in a net magnetization of $\rm{10 \mu_B}$ for the low temperature limit \cite{CHEREPANOV199381}.  
Its excellent magnetic properties such as very low intrinsic  damping factor of $\rm{10^{-3}}$ to $\rm{10^{-5}}$ \cite{Kajiwara_2010,Heinrich_2011,Kurebayashi_2011} and high spin density of $\rm{ 2.1 \cdot 10^{22}\,\mu_B\, \text{cm}^{-3}}$ per unit cell \cite{Gilleo_1958} render YIG a widely used material in strong coupling experiments between microwave photons and magnons \citep{Tabuchi_2014,Zhang_2014,Haidar_2015,Klingler_2016}. 
The YIG sphere is glued to a ceramic (BeO) rod along the $\rm{[110]}$ crystal direction. The saturation magnetization $M_s$ has been determined via SQUID measurements along the same axis to $\mu_0  M_{\rm{s}}=(282 \,\pm\, 3)\,\rm{mT}$.\par
\begin{figure}[t]
\includegraphics[scale=0.35]{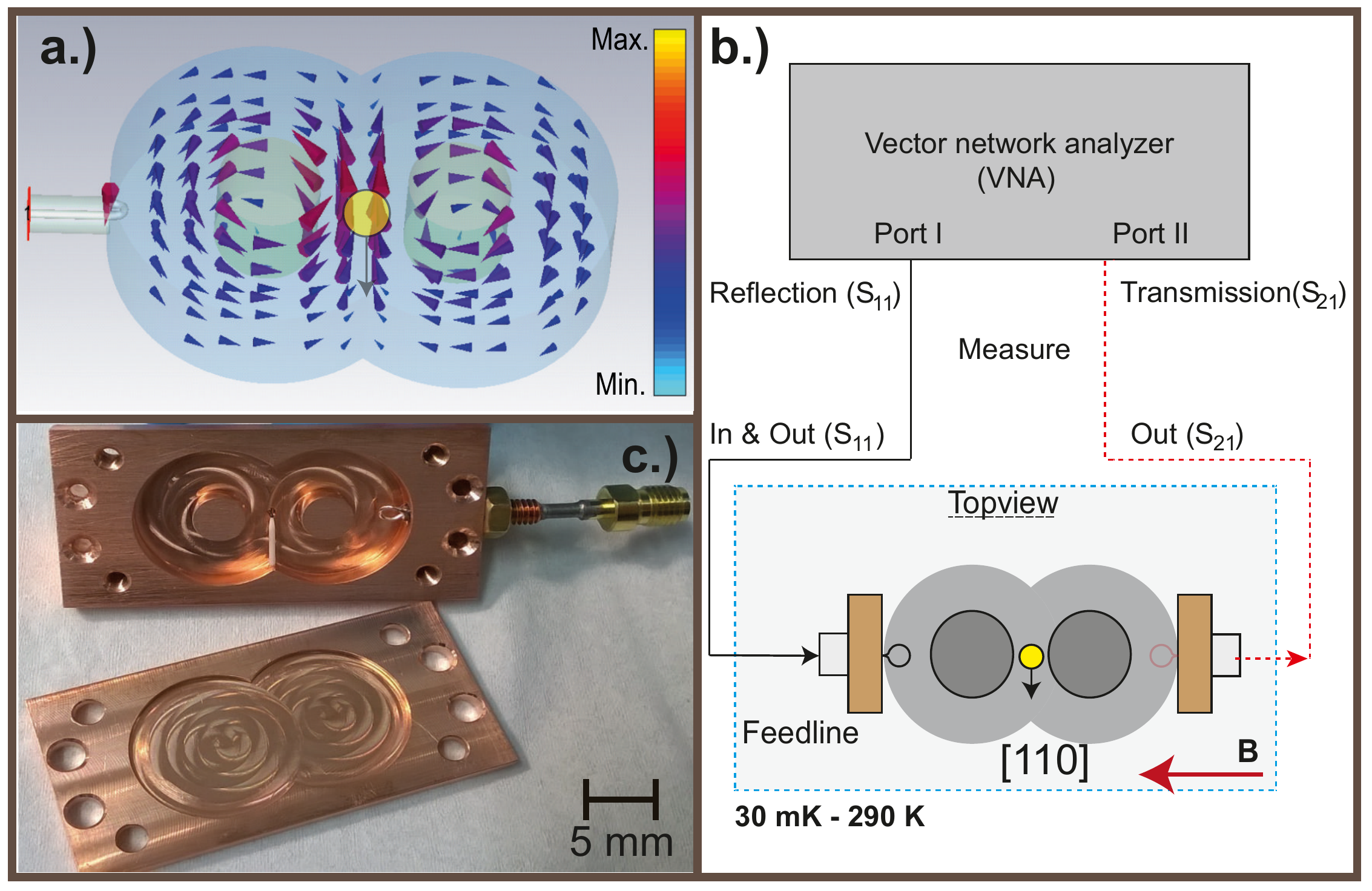}
\caption{ Experimental setup and position of  the YIG sphere in cavity resonator. a.) Simulated magnetic field distribution of the `bright´ mode resulting in an enhancement between the posts at resonance b.) Schematics of the experimental setup. For reflection  and transmission  measurements one or two loops with one winding are used for inductively coupling the microwave signal from the VNA into the resonator. The YIG sphere is placed in the magnetic field's antinode in the middle between the posts for enhanced coupling in the bright mode. c.) Photograph of the assembled YIG-cavity resonator system. }
\label{Setup}
\end{figure}
\begin{figure}
\includegraphics[scale=0.4]{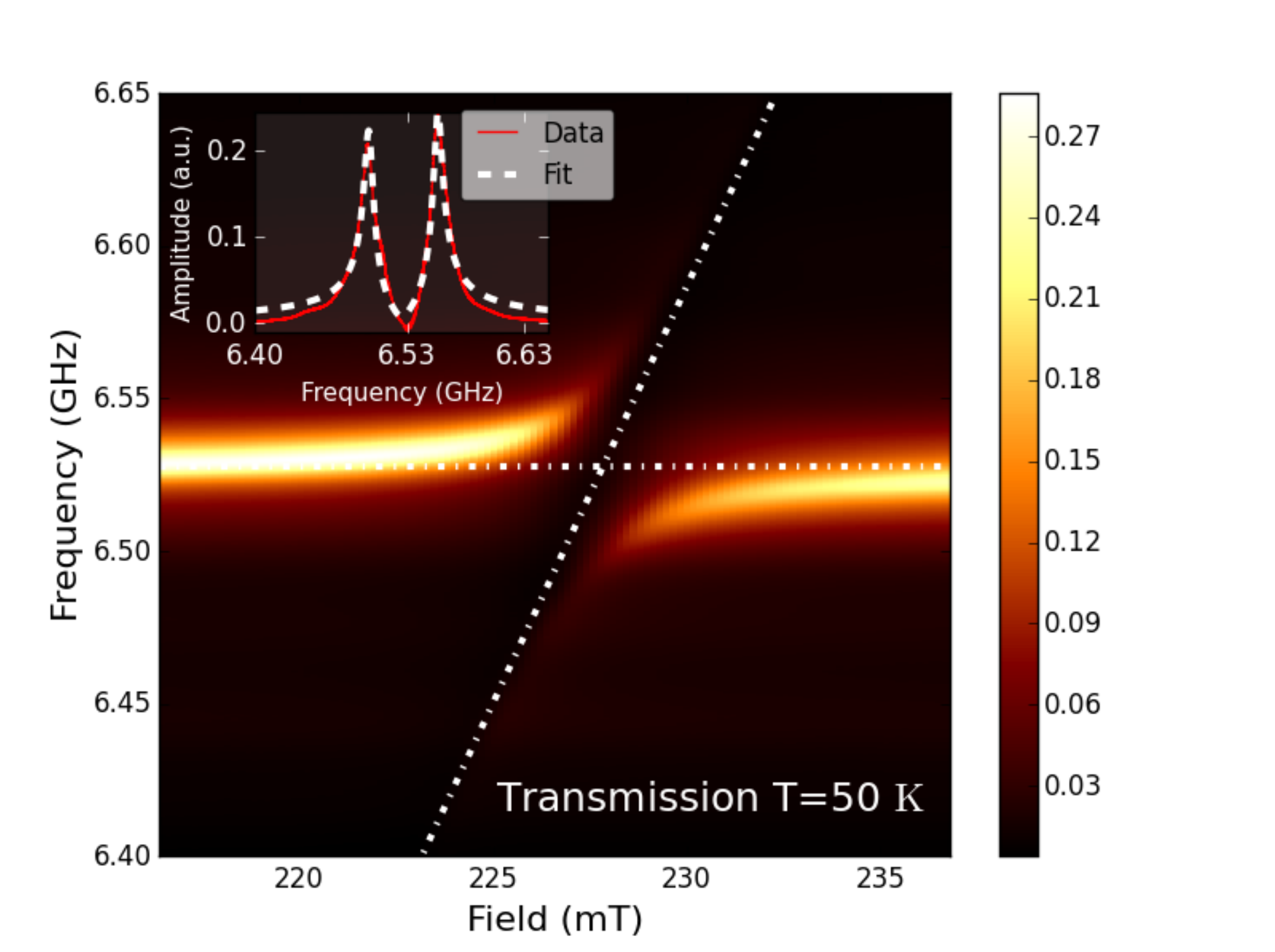}
\caption{Transmission ($\mathcal{S}_{\rm{21}}$) measurement of the dispersion for the cavity-magnon polariton for $T=50 \,\rm{K}$. Resonant coupling appears at $\rm{227\,mT}$ with a coupling strength of $g_{\rm{eff}}(T)/2\pi=29\,\rm{MHz}$. The raw data (not shown) had magnetic field independent oscillations due to standing waves in the cables, for analysis of the data they were removed as described in the supplement. As an example, the inset shows the background corrected resonant linear amplitudes of the coupled system with a fit of Eq. \ref{S21} to the data. The red line shows the background corrected data, the other one (dotted) the corresponding fit. The dotted lines in the main plot illustrate the uncoupled dispersion of the cavity (field independent) and Kittel magnon (field dependent).}
\label{dispersion}
\end{figure}

For our resonator we use a reentrant cavity resonator design with an enhancement of the  magnetic field density in the center, see Fig. \ref{Setup} a.). This is realized by a cavity resonator with two cylindrical posts in the cavity. 
Depending on the direction of current of each post, the field between the posts results in `bright'  and `dark' modes \cite{Goryachev_2014}. In the `bright' mode the post's magnetic fields interfere constructively and enhance the total magnetic field strength between the posts. Destructive interference leads to a vanishing magnetic field in the `dark' mode.  
Our cavity resonates at $5.5\,\mathrm{GHz}$ in the `dark´ and at $6.5\,\mathrm{GHz}$ in the `bright´  mode at room temperature. From simulations we found that the next cavity mode resonates at frequencies of $18\,\mathrm{GHz}$, thus the coupling of  microwave energy into other cavity resonator  modes than the bright one is strongly suppressed. 
The microwave signal is provided by a vector network analyzer (VNA) with an output power level of $P=0\,\mathrm{dBm}  (1 \,\mathrm{mW})$ for all measurements from $T=5\,\mathrm{K}$ to $T=290\,\mathrm{K}$. The total distance (cable length) from the VNA to the sample and vice versa  is $\approx 5\,\textrm{m}$ with an attenuation of $3\, \textrm{dB}$ per meter. Correspondingly, this yields an attenuation of $(7.5 \pm 0.5)\,\textrm{dBm}$ with taking uncertainties in the absolute cable length and the changes due to different temperatures into account .
From a circle fit to the data of the reflection cavity, the calculation for the loaded quality factor $Q_{\rm{l}}$ yields $Q_{\rm{l}}=(1804 \pm 13) $ and  $Q_{\rm{e\\
}}=(4074 \pm 22) $ at $T=290\,\mathrm{K}$ \cite{pozar}.
The average photon number in our cavity resonator $\braket{n}$ yields $\braket{n}=(3.23\pm 0.74)\cdot 10^{12}$  with the attenuation given before due to the influcence of the signal line at the cavity resonator. 
The feedline signal is coupled into the cavity resonator via a single winded loop (cf.  Fig.\ref{Setup}). Rotation of the loop within the cavity resonator tunes the coupling strength between feedline and cavity resonator. Reflection and transmission cavities have been employed, see Fig. \ref{Setup}. Specifically, we measured with two different cavities from an identical design with the only difference being the number of coupling loops. This means, we performed reflection measurements only  with a single port (one coupling loop) cavity resonator and transmission and reflection measurements with a cavity resonator with two ports (two coupling loops).  Slight differences in production, mounting or filling factors cause different resonant fields for otherwise identical conditions (cf. Fig. \ref{dispersion} and Fig. \ref{Fig6_kappa_resonance}). For the temperature dependent measurements from $5\,\mathrm{K}$ to $290\,\mathrm{K}$, we use a ${}^4 \mathrm{He}$  continuous flow cryostat with a superconducting magnet at the bottom. 
The milliKelvin data (measured in reflection only) is obtained in a dry dilution refrigerator setup with $110\,\rm{dB}$ attenuation on the incoming line, a circulator in front of the sample and cryogenic amplifier at $4\,\rm{K}$ on the return line. The output power level of the VNA was at $-25\,\rm{dBm}$, which results in a total power at the coupler of $-135\,\rm{dBm}$. This corresponds to $\braket{n}< 1$ photons in the cavity resonator, thus we enter the single photon regime where the quantum effects come into play \cite{Tabuchi2015}.  We probe our system using ferromagnetic resonance techniques with a VNA and standard tools of network analysis. For data aquisition and analysis we use the open-source toolbox \textit{qkit} \cite{QKIT,probst}.

The microwave scattering parameters of the  magnon-photon system were measured from $290\,\mathrm{K}$ down to $30\,\mathrm{mK}$. For each temperature we obtained a spectrum such as shown in Fig. \ref{dispersion} for $T=50 \, \mathrm{K}$. In a first step, we fit the dispersion of the coupled branches from cavity resonator and magnons in Kittel mode ($\vec{\textbf{k}}=0$) in the YIG sphere and extract the coupling strength $g_{\rm{eff}}(T)$ as function of temperature.\\
By means of Eq. \ref{Eq:coupling_strength}, we first compare the theoretical expectation of the single spin coupling strength $g_0^{\rm{th}}$ with our experimental findings for the coupling strength at room temperature. 
Numerical simulations yielded a cavity mode volume  $V_{\rm{mode}}=9.53 \cdot 10^{-7}\, \mathrm{m^3}$, leading to 
$g_0^{\rm{th}}/2\pi=\rm{18.82 \, \mathrm{mHz}}$ for $\eta=1$ \cite{Denner}.  
Computing the field gradient over the volume of the YIG sphere  we obtain an estimated value of $\eta=0.58$ \citep{Denner}. Thus, we expect $g_0/2\pi=10.92\,{\rm{mHz}}$ and an effective coupling strength $g_{\rm{eff}}=g_{\rm{th}} \sqrt{2Ns}$. A YIG sphere of a radius of $d=(0.25\,\pm\, 0.01)\,\mathrm{mm}$  taking deviations into account, yields $N=(1.374 \,\pm\, 0.15)\cdot 10^{18}$ spins, thus we compute $g_{\rm{eff}}/2\pi =(28.77\,\pm\, 3.29) \,\mathrm{MHz}$ as an upper value. 
\begin{figure}[t]
\centering
\includegraphics[scale=.3]{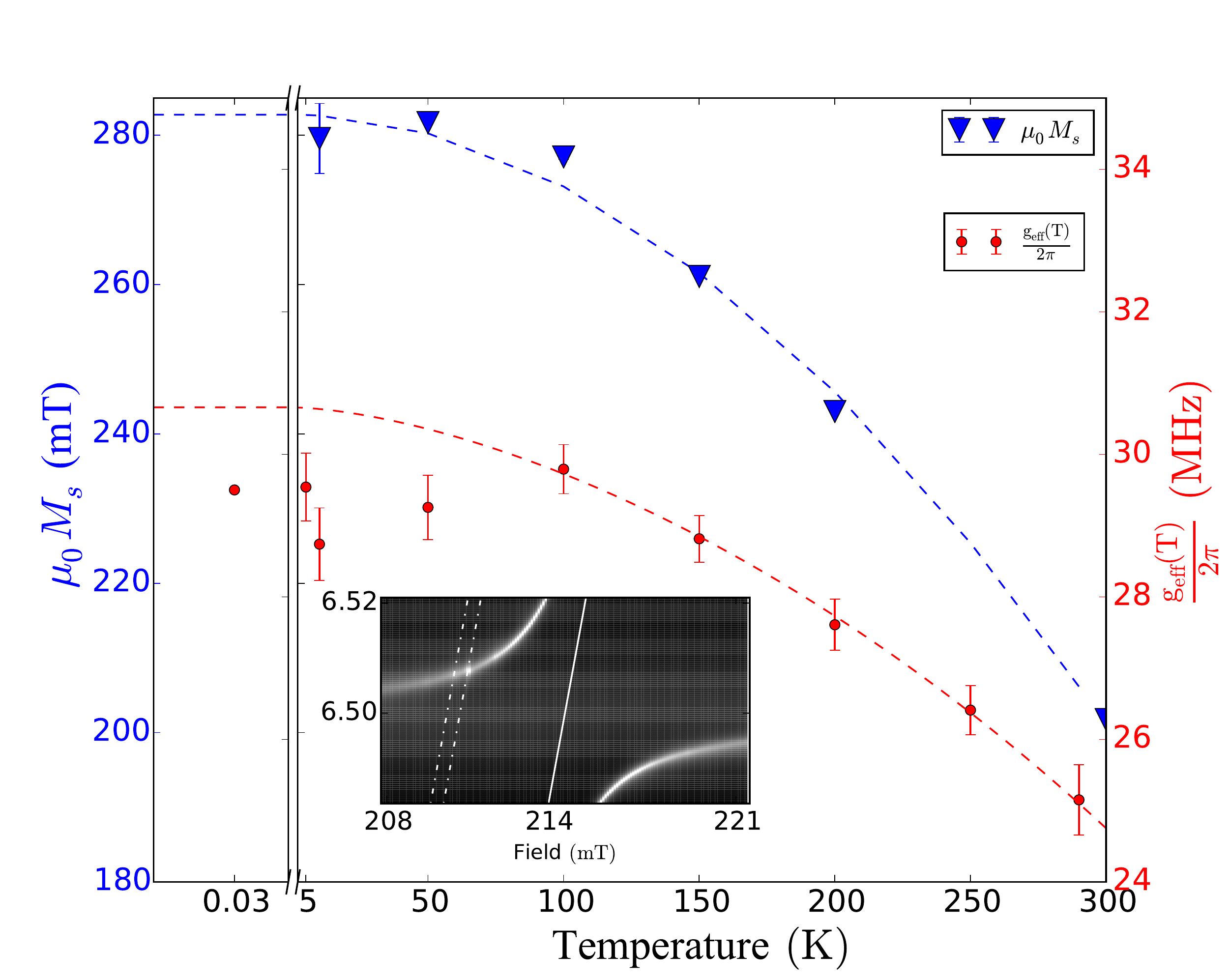}
\caption{Temperature dependence of the saturation magnetization $\mu_0 M_{\rm{s}}$ (SQUID measurements) and the effective coupling strength $g_{\rm{eff}}(T)$ of the Kittel mode measured with the single port cavity resonator (reflection). The dotted lines in the main figure correspond to fits of the data to Bloch's $T^{3/2}$ law. The left inset shows a zoom into the region around resonance of the magnon polariton for a temperature of 100 K. The dotted lines indicate additional couplings to parasitic spin wave modes close to the Kittel mode. For lower temperatures the fields for the couplings start overlapping and the coupling strength of the Kittel mode decreases. The left shows exemplarily a value of the subKelvin temperature measurement which is consistent to the higher temperature data ($T \geq 5\,\mathrm{K}$). The milliKelvin regime is discussed in detail in a separate publication \citep{pfirr}.}
\label{gMs}
\end{figure}
In the main plot of Fig. \ref{gMs}, we show for $g_{\rm{eff}}(T)$ along with a measurement of the saturation magnetization $M_{\rm{s}}$ of our YIG sphere. 
We identify two regimes for the temperature dependence of $g_{\rm{eff}}(T)$:  $T< 100\,\mathrm{K}$ and  $T\geq 100\,\mathrm{K}$. \\
We first consider the $T\geq 100\,\mathrm{K}$ regime. There, both the coupling strength $g_{\rm{eff}}(T)$ and the saturation magnetization $M_s(T)$ increase in a similar fashion. $M_s(T)$  is proportional to the number of spins $N$ and changes with temperature according to Bloch's $T^{3/2}$ law; $M_{\rm{s}}(T)= M_{\rm{s}}(0)\left[1-\left(\frac{T}{T_{\rm{c}}}\right)^{\frac{3}{2}}\right]$, where $M_{\rm{s}}(0)=\frac{\rho}{\mu_0 }\mu_{\rm{B}}$ with a net spin density $\rho$  \cite{gross}.
We can link the saturation magnetization $M_{\mathrm{s}}(T) \propto N$ and the coupling strength $g_{\rm{eff}}(T)$ using $g_{\rm{eff}}(T)\propto \sqrt{N}$ (Eq.\ref{Eq:coupling_strength}). Thus, valuable insight on the participating spins is provided by the relation  $g_{\rm{eff}}(T)/ \sqrt{M_{\rm{s}}(T)}$.  
If the change in $N$ would be the dominant contribution to the changes of $g_{\rm{eff}}(T)$ (shown in Fig. \ref{Fig_8_ratio} of supplementary information), it should not change its value significantly. 
In principle, changes in physical properties besides $M_{\rm{s}}$, such as $\omega_{\rm{r}}$ or $\gamma$, could account for the observed temperature dependence of $g_{\rm{eff}}(T)$ as well. The resonance frequency of the bright mode $\omega_{\rm{r}}/2\pi$ changes less than $\rm{0.5\,\%}$ (cf. Fig. \ref{res_1} b.)). The gyromagnetic ratio $\gamma$ calculated from the ratio between resonance frequency $\omega_{\rm{m}}$ and field $H_{\rm{res}}$ changes only by at most $\rm{4\,\%}$ (cf. Fig. \ref{res_1} b.)) over the whole temperature range due to the change in the resonance field. \cite{Kittel_1948}. However, $M_{\rm{s}}(T)$ changes by $\approx 27\,\%$, and we conclude that $N$ has the strongest contribution on the temperature behaviour of ${g_{\rm{eff}}(T)}$ above 100 K. The dotted lines in Fig. \ref{gMs} show a  fit of the data to Bloch's law in first order, which reads to 
$M_{\rm{s}}(T)=M_{\rm{s}}(0)\left[1-\left(\frac{T}{T_{\rm{c}}}\right)^{\beta \cdot\frac{3}{2}}\right]$  and $g_{\rm{eff}}(T)=g_{\rm{eff}}(0)\sqrt{\left[1-\left(\frac{T}{T_{\rm{c}}}\right)^{\beta \cdot\frac{3}{2}}\right]}$, where $\beta$ is a fitting parameter, respectively.  This behaviour for $g_{\rm{eff}}(T)$ is only valid for $T\geq 100\,\mathrm{K}$, thus lower temperature data was not included for the fit. 
Fitting $g_{\rm{eff}}(T)$ yields
$g_{\rm{eff}}(0)/2\pi=(30.65 \,\pm\, 0.5) \, \mathrm{MHz}$, $T_{\rm{c}}=(579 \,\pm\, 43)\, \mathrm{K}$ and $\beta=1.06 \,\pm\, 0.15$. Accordingly, the results for $M_{\rm{s}}(T)$  read $\mu_0 M_{\rm{s}}(0)=(282 \,\pm\, 3)\,\rm{mT}$, $T_{\rm{c}}=(566 \,\pm\, 43)\, \mathrm{K}$  and $\beta=(1.30 \,\pm\, 0.15)$. 
From the fit we obtain an upper value for  $g_{\rm{eff}}(T)/2\pi= 30.65\,\mathrm{MHz}$ for decreasing temperatures, which is slightly higher than the calculated value of $28.77\,\mathrm{MHz}$. From the spin net density from literature, $\rm{ 2.1 \cdot 10^{22}\,\mu_{\rm{B}}\, \text{cm}^{-3}}$ \cite{Gilleo_1958}, we calculate for $M_{\rm{s}}(T=0\, \mathrm{K})=194 \,\pm\, 32 \frac{\mathrm{kA}}{\rm{m}}\; (\mu_0 M_{\rm{s}}=(245 \,\pm\, 40)\,\mathrm{mT})$. Within the errrorbars this is in line with  the  fitted value determined from SQUID measurements of $M_{\rm{s}}=(282\,\pm\,3)\,\mathrm{mT}$. Though, the absolute value is higher than the value reported in literature \cite{gross}. This  difference can be caused by an uncertainty in $N$. Assuming a deviation of the sphere's radius of $0.01\,\mathrm{mm}$, the uncertainty of $\mu_0 M_{\rm{s}}$ is $\,\pm\, 40 \,\mathrm{mT}. $ Therefore, only by taking such deviations into account, the measured and calculated value are in accordance.  Additionally, an inhomogeneous magnetic field between the two posts of our specific cavity design affects the measured coupling strength. In our design, the maximal field is located close to the posts and decreases to the central position between them where the YIG sphere is placed. Field variations over the sphere and some misalignment with respect to the center lead to changes in $\eta$.\\ 
While $g_{\rm{eff}}(T)$ for $T>100\,\mathrm{K}$ is well correlated to the temperature dependence of the magnetization, for $T < 100\, \mathrm{K}$ the coupling strength $g_{\rm{eff}}(T)$ behaviour is different. As shown in the left inset of Fig. \ref{gMs} additional anticrossings of much smaller coupling strength appear at lower values of the applied external field but still rather close to  the resonance field of the Kittel mode.
For even lower temperatures, the resonant fields for both Kittel mode and higher order modes shift to lower values (for details see supplementary information). However, the resonant field of the Kittel mode shifts more strongly \citep{Gennes_1959,Kittel_1948}. Eventually, the higher order modes become degenerate with the Kittel mode. In this case, the probability for cavity photons to couple to more modes besides the Kittel mode is increased. As the number of total spins $N$ is conserved, we observe a decrease in the coupling strength $g_{\rm{eff}}(T)$ measured in the anticrossing of the Kittel mode, see  Fig. \ref{gMs}.
When the temperature is lowered further, the coupling strength $g_{\rm{eff}}(T)$ starts increasing towards the saturation value. As shown in the top right inset of Fig. \ref{gMs}, we measure  $g_{\rm{eff}}(T)/2\pi=29.5\, \mathrm{MHz}$ at $T=30\,\mathrm{mK}$, which is in the same range as both the calculated $g_{\rm{0}}^{\rm{th}}$ and the fitted $g_{\rm{eff}}(0)$ values.\\
The cavity-magnon polariton is the associated quasiparticle of a coherent interaction  between strongly coupled magnons and (cavity resonator) photons. Losses from each subsystem could in principle affect the magnon polariton and have to be taken into account. 
Thus,  for a full temperature dependent picture  of the magnon polariton it is not sufficient to examine the temperature evolution of the coupling strength only but to study the temperature dependence of the system's  dissipation and specifically the  magnon linewidth $\kappa_{\rm{m}}(T)$, as well. 
This combination of ascertaining both coupling and dissipative losses allows then for the key insights of the coupling regime over the full temperature range. Accordingly, we can then quantify our coupling regime by including dissipation from both the magnons and the resonator  by the dimensionless cooperativity parameter $C=\frac{g_{\rm{eff}}^2(T)}{\kappa_{\rm{m}}(T) \kappa_{\rm{l}}(T)}$. The strong coupling regime is reached for values of C greater than one \cite{9783540285731}. \\
The magnon linewidth is comprised of contributions of homogeneous and inhomogenous processes from two magnon scattering which results in $\kappa_{\rm{m}}/2\pi=\alpha \frac{\omega}{2\pi}+ \frac{\Delta \omega}{2\pi} $, where $\alpha$ denotes the phenomenological Gilbert damping parameter with a linear dependence on the frequency $\omega$ and $ \Delta \omega/2\pi$ the contribution from inhomogenous processes. 
If multiple scattering processes are present at the same time, $\Delta \omega$ denotes the sum of single linewidth contributions and $\Delta \omega=\sum_i \Delta \omega_{i}$. For YIG, the Gilbert damping parameter  with a very low value of $\alpha$ can increase slightly as a function of temperature \cite{Haidar_2015}. However, the observed values for the linewidth decrease towards higher temperatures, and contributions from scattering processes are dominanting the temperature dependence \cite{Klingler_2016}. 
In general, spin-spin interaction processes can lead to a redistribution of the energy in the magnon mode spectrum. Among them, one distinguishes between intrinsic and extrinsic processes. The first occur in perfect crystal structures and the latter describe processes due to scattering off defects \cite{MelkovBook}. In the presence of inhomogeneities from geometrical imperfections and impurities within the magnonic specimen \cite{Hurben_1998}, surface pit and rare earth impurity scattering, respectively, have to be taken into account.\\
The intrinsic relaxation of  the Kittel mode is described by the Kasuya-Le Craw mechanism and is for highly pure YIG crystals typically in the order of $0.01\, \mathrm{mT}$, or a frequency linewidth of $0.28 \,\mathrm{MHz/2\pi}$ \cite{kasuya,MelkovBook}.  
In principle, the two beforehand mentioned extrinsic processes could make significant contributions: First, geometric imperfections such as pores or rough surfaces result in spatially non-uniform samples and thus different conditions for ferromagnetic resonance employed in the Kittel mode. This broadens the linewidth as the resonance frequency is non-uniform throughout the sample. Second, scattering at rare earth impurity ions with a strong spin-orbit coupling within the YIG specimen can contribute strongly to the linewidth as well. The linewidth peaks when the temperature dependent energy splitting of the ion's groundstate multiplet approaches the microwave photon energy in the chosen cavity resonator mode \cite{dillon}.\\
We extract the values for the magnon linewidth $\rm{\kappa_{\rm{m}}}$ from a fit of our background corrected reflection (transmission) amplitude data taken with a cavity resonator with two ports to input-output theory via Eq.  \ref{S11} and \ref{S21}, respectively. Further information on our background correction scheme and the fitting procedure can be found in the supplement. 
The temperature dependence of the linewidth $\kappa_{\rm{m}}/2\pi$ of the Kittel mode at resonance is plotted in Fig. \ref{km}, where the circles (red) refer to the reflection and the triangles (blue) to the transmission data, respectively. Within the error bars, the temperature dependence for the reflection and transmission measurements reveal the same temperature dependence of the linewidth $\kappa_{\rm{m}}/2\pi$, as has been expected.  
For subKelvin temperatures (measured in reflection only), as shown in the inset of Fig. \ref{km}, the measured linewidth $\kappa_{\rm{m}}/2\pi=1.2 \,\text{MHz}$ is in good agreement to the literature value as reported in Ref. \onlinecite{Tabuchi_2014}. The details of the behaviour in the subKelvin temperature regime are beyond the scope of this work and will be the topic of a future study \cite{pfirr}. Starting from the lowest temperature, we observe an increase of the magnon linewidth with increasing temperature. As shown in Fig. \ref{km} one dominant peak of the magnon linewidth appears near $40\,$K. Additionally, the curve's modeling yields another, relatively broad peaklike structure around  $100\,$K. 
Similar temperature behaviour has been observed for LPE grown YIG films with X-Band FMR measurements \cite{vittoria}, in line with our measurement with a bulk YIG sample. 
\begin{figure}[t]
\centering
\includegraphics[scale=0.3]{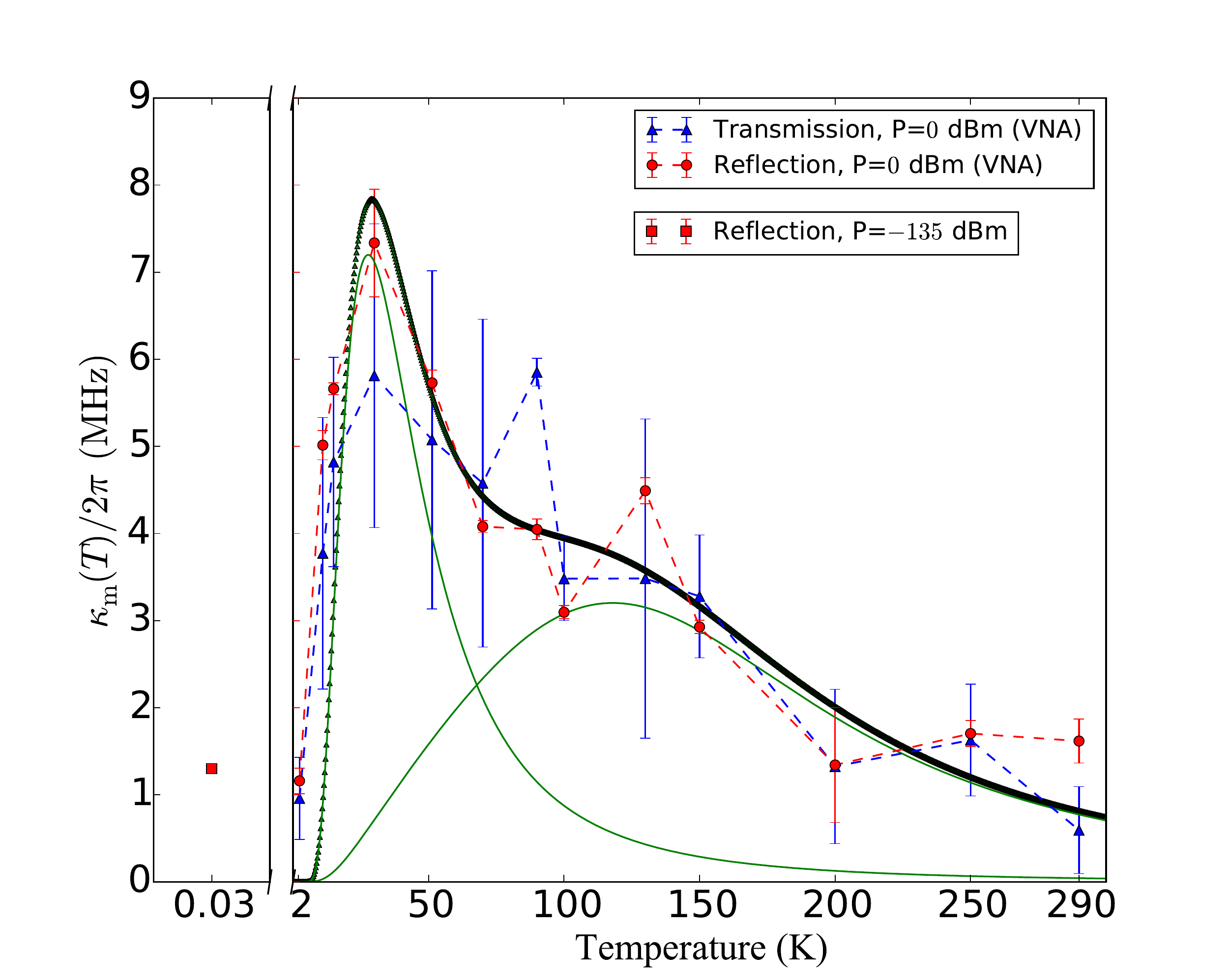}
\caption{ Magnon linewidth $\kappa_{\rm{m}}$ as derived from a fit to Input-Output theory via Eq. \ref{S11} and \ref{S21} for reflection (red, circles)  and transmission (blue, triangles) data (measured with our cavity resonator equipped with two ports). The error bars are obtained from the fit's covariance matrix. Higher error bars for the transmission data result from the  influence of additional avoided crossings in the vicinity of the main splitting. The data taken in the milliKelvin setup with $-135\,\rm{dBm}$ input power at the inductive loop coupler of the cavity resonator is shown on the left side of the figure. We observe a total change of the magnon linewidth by a factor of six in the range between $30\,\mathrm{mK}$ to $290\,\mathrm{K}$ with a peak at $43\, \mathrm{K}$. This behaviour is goverened by impurity rare earth ion relaxation  \cite{Spencer_1961,Seiden_1964,Klingler_2016,Jermain_2017}. The green curve models the temperature dependence, taking a possible contribution from rare earth impurity scattering at different temperatures into account. The sum of them results in the black line (stars). The milliKelvin data is analyzed closer in a separate publication \cite{pfirr}. }
\label{km}
\end{figure}
To understand this behaviour, following Ref. \cite{Gennes_1959,Sparks1961}, scattering due to rough surface polish  is modeled by small semispherical pits  with radii of $2/3$ of the surface quality of the polishing material. The expected linewidth is proportional to the saturation magnetization, thus the same temperature dependence is expected with the largest contribution to the linewidth occurring at the lowest temperatures.
This is expressed by: 
\begin{equation}
\frac{\Delta \omega_{\rm{surface}}(T)}{2\pi}\propto \gamma \frac{R_{\rm{pit}}}{R_{\rm{sample}}} 4\pi M_{\rm{s}}(T),
\label{rough}
\end{equation} 
where $R_{\rm{pit}}$ denotes the average radius of the surface pits, and $R_{\rm{sample}}$ the radius of the used spherical sample. An estimate using Eq. \ref {rough} yields a contribution to the linewidth in the order of $2\pi \cdot 1\,\mathrm{MHz}$. This background contribution is too small for the observed linewidth and cannot explain the observed peaks.
The relaxation due to rare earth impurities dominates and leads to the observed peaking at low temperatures.
 The temperature dependence of this linewidth is described by: 
\begin{equation}
\frac{\Delta \omega_{\rm{REI}}(T)}{2\pi} \propto \omega \frac{e^{\hbar\Omega_{\rm{0}}/k_{\rm{B}}T}}{[e^{\hbar\Omega_{\rm{0}}/k_{\rm{B}}T}+1]^2}\frac{\tau_{\rm{imp}}}{\tau^2_{\rm{imp}}+\omega_{\rm{m}}^2},
\label{REI}
\end{equation}
where $\hbar\Omega_{\rm{0}}/k_{\rm{B}}$ denotes the temperature where the peak of the magnon linewidth based on rare earth impurity scattering is measured and $\rm{\tau_{imp.}} $  represents the impurity relaxation rate. Depending on the specific rare earth element, peaks occur up to temperatures of  $150\,\mathrm{K}$ \cite{Seiden_1964,safo}. In our experiment, we find one peaking at $43\,\mathrm{K}$ and a broad small increase around $100\,\mathrm{K}$. An exact determination of the specific rare earth elements requires further investigations which is beyond the scope of this paper. In Fig. \ref{km}, we used this relation to model the temperature behavior of our magnon linewidth (green curve) with the dominant contribution appearing at $\rm{\hbar\Omega_0/k_B=43\,K}$ with a linewidth of $7\,\mathrm{MHz}$. Another second contribution was estimated to be at $90 \,\mathrm{K}$ with a linewidth of $3.5\,\mathrm{MHz}$. The relaxation rate of the spins to the lattice is modeled with $\mathrm{\tau_{imp}=\zeta} T^2$, with $\zeta/2\pi \rm{ =0.2 \cdot 10^{-3}}\,\mathrm{GHz/K^2}$ for the first and $\zeta/2\pi \rm{=29\cdot 10^{-3}}\,\mathrm{GHz/K^2}$ for the second peak, respectively. 
Approaching room temperature the linewidth decreases to a value of $\rm{\kappa_{\rm{m}}/2\pi=1.2\,}$MHz at room temperature. This linewidth agrees with previous values reported in literature \citep{Zhang_2014}. 
\begin{figure}[h!]
\centering
\includegraphics[scale=0.3]{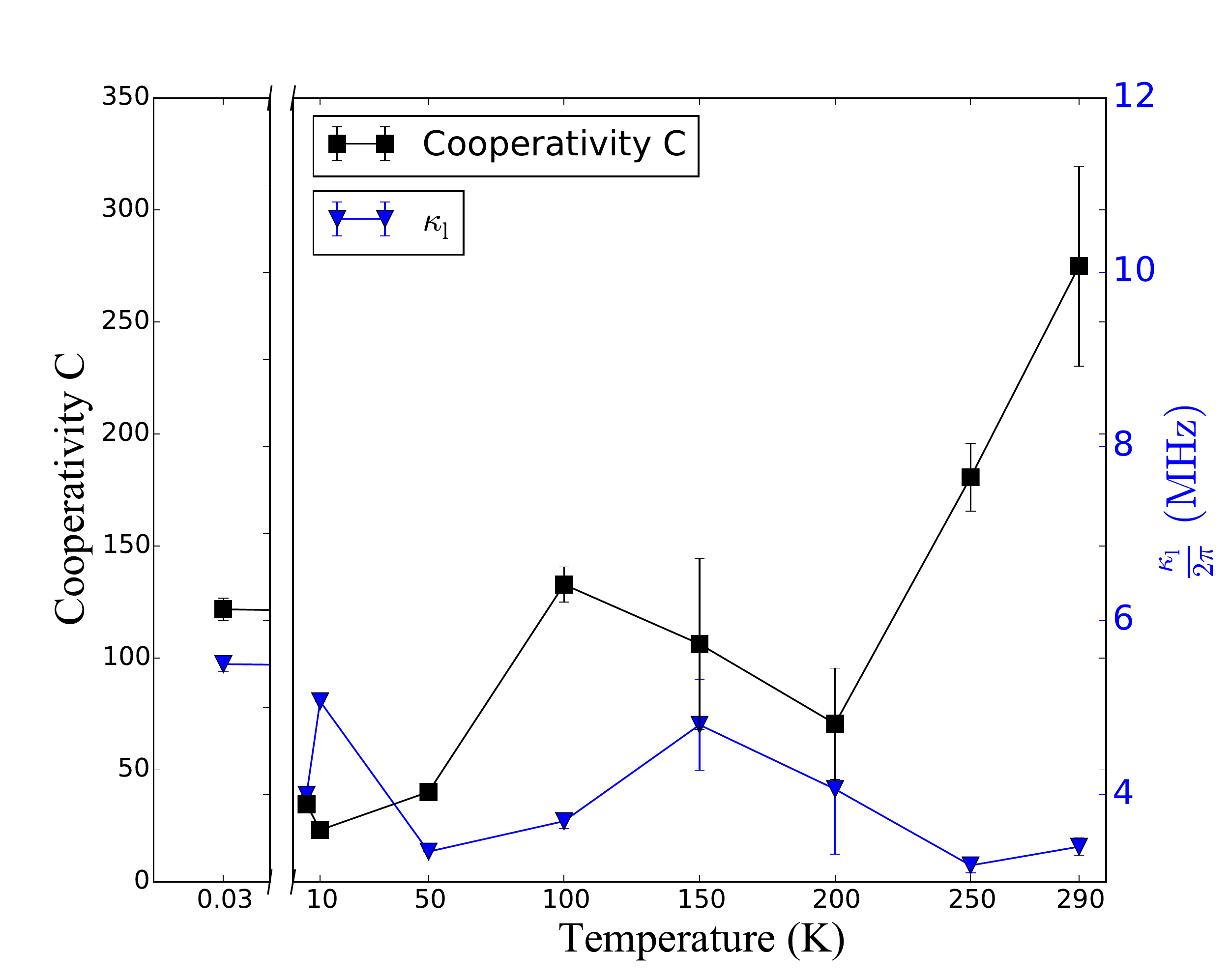}
\caption{ Temperature dependence of the cooperativity $C$ (squares) and  the total cavity resonator losses $\kappa_{\rm{l}}(T)$ (triangles)  from the measurement with the single port cavity resonator. The cooperativity decreases with decreasing temperature but is much larger than one for all temperatures. In contrast to the magnon linewidth $\kappa_{\rm{m}}$ the total cavity resonator loss $\kappa_{\rm{l}}$ do not change significantly as a function of temperature.
Thus, the temperature dependence of the cooperativity inversely reflects the changes of the magnon linewidth. }
\label{Cooperativity}
\end{figure}
Finally, we calculate the electromagnonic cooperativity $C$, as displayed in Fig. \ref{Cooperativity}.
The cooperativity is much larger than one and it shows, that the coupling strength exceeds the losses for all temperatures in this study. Over the entire temperature regime the cooperativity exceeds $20$, with values beyond $100$ in the milliKelvin regime as well as at room temperature. 
Accordingly, such large cooperativity values over the entire temperature range firmly place our hybrid system in the strong coupling regime, where coherent dynamics between both subsystems occur. Robust coherence is one of the key properties for future applications in information technology \cite{Divincenzo2005}. Thus, such a hybrid system can be a suitable candidate since it can combine the advantages of each subsystem for further use. \\  
To conclude, in order to connect the cryogenic and room temperature regimes we performed temperature dependent measurements on a hybridized microwave cavity magnon system. At resonance, a quasiparticle, the cavity-magnon polariton is formed and observed as an anticrossing in the spectrum. Correspondingly, we studied the temperature dependence of the coupling strength $g_{\rm{eff}}(T)$ and the linewidth $\kappa_{\rm{m}}$ and linked both quantities via the cooperativity C showing strong coupling prevails for all temperatures.  For temperatures $T>100\,K$ the coupling strength is governed by Bloch's law due to the scaling with the number of spins. For lower temperatures an additional coupling of the microwave photons to other modes in the sphere leads to a decrease in $g_{\rm{eff}}(T)$, because the total number of spins is conserved. We find that for such coupled system, the temperature dependence of the linewidth $\kappa_{\rm{m}}$ is dominated by rare earth impurity scattering. These measurements show the coupling and dissipation of cavity-magnon polaritons for instance at transitions between the single photon and the classical regime. In particular, we show that spectroscopy on strongly coupled magnon-photon systems is a versatile tool for spin material studies over large temperature ranges, providing key parameters with the same measurement. 
We acknowledge valuable discussions with Hannes Rotzinger, Carsten Dubs and Michael Denner. This work was supported by the European Research Council (ERC) under the Grant Agreement 648011 and through SFB TRR 173/Spin+X.

\section*{Supplementary Material}
\subsection*{\label{sec:Dataanalysis}{Methods}}
The coupling strength $g_{\rm{eff}}(T)$ and the magnon linewidth $\kappa_{\rm{m}}$ were extracted from fitting our data to the complex scattering parameters for reflection and transmission (equations \ref{S11} and \ref{S21}). The following steps are done to remove background resonances before the fitting. 
\begin{itemize}
\item[1.] The cables from the VNA to the cavity-YIG system positioned at the end of the measurement setup contribute to the measured signal with a constant, i.e field independent background, in form of a standing wave. These signals reduce the signal-to-noise-ratio such that, for instance, at $5\,\mathrm{K}$ the signal is at the noise level for a length $5\,\mathrm{m}$ of the signal line ($7.5 \,\mathrm{dB}$). Therefore,  the data needs to be corrected for that field modulation. We perform a weighted average over all field points for each entry in the frequency window.  By assigning zero weight to a selected area around both the cavity and the Kittel magnon dispersions the background is averaged over all fields. That average is then removed from the scattering amplitude.
\item[2.] Secondly, we perform a fit based on the standard least-squares algorithm to the avoided crossing in our data by solving the $2 \times 2$ matrix of the energy eigenmodes of a system of two coupled harmonic oscillators at resonance. Further, we do a circle fit to the cavity far away from resonance \citep{QKIT}. This yields the coupling strength $g_{\rm{eff}}(T)$, the resonance frequency $\omega_{\rm{r}}$ of the cavity resonator, the ferromagnetic resonance frequency of the magnon $\omega_{\rm{m}}$ and the quality factors $Q_{\rm{i}}$ and $Q_{\rm{e}}$, i.e. loss parameters of the cavity, respectively. The results from these fits are used as input parameters for the final fit of the scattering function. 
\item[3.] Finally, the scattering amplitudes  are fit according to equations Eq. \ref{S11} and \ref{S21} to the background corrected data and the relevant parameters are extracted. This fit is based on the least-squares alogrithm as well.
\end{itemize}
\subsection*{Temperature dependence of other quantities affecting the coupling strength $g_{\rm{eff}}(T)$  }
\begin{figure}[h!]
\centering
\includegraphics[scale=0.3]{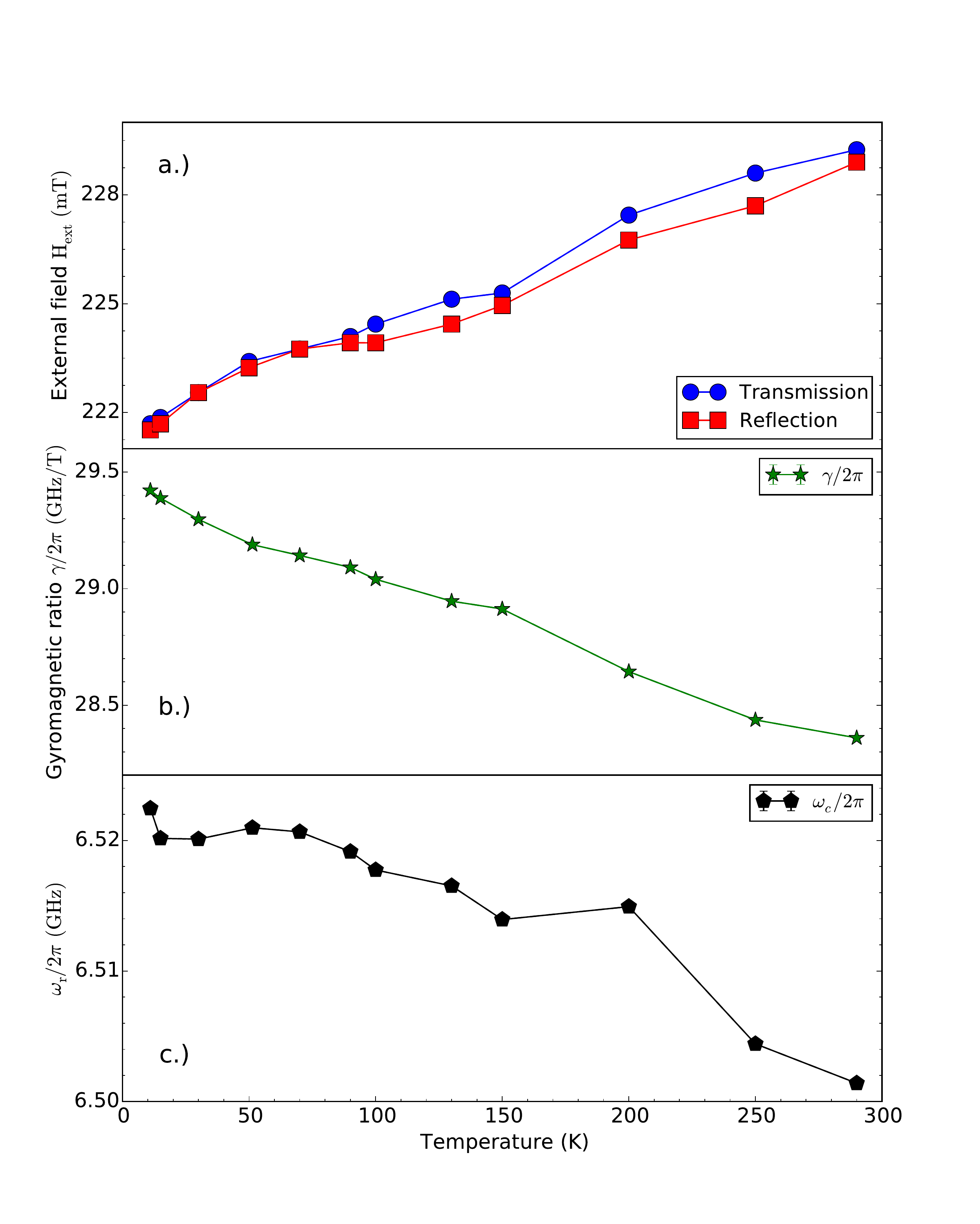}
\caption{ a.) Temperature dependence of the applied static field in the resonance condition for the Kittel mode. The external static field values for resonant coupling increases by $4\,\%$ from $5\,$ to $290\, \mathrm{K}$. The increase is attributed to the decrease of the saturation magnetization $M_{\rm{s}}$ as a function of temperature (cf.  Fig. \ref{gMs}) since demagnetizing fields have to be taken 
into account, as well. Thus, $H_{\rm{res}}=H_{\rm{ext}}-H_{\rm{demag}}$. The error bars are in the order of $0.1\,\mathrm{mT}$, and are covered by the data symbols.
b.) Temperature dependence of the gyromagnetic ratio. The values are calculated from $\rm{\omega_{res}=\gamma\cdot H_{ext}}$. It decreases towards higher temperatures due to the changes in $H_{\rm{demag}}$. The total decrease is less than $4\,\%$ and thus small compared to the temperature dependence of the coupling strength $g_{\rm{eff}}(T)$. 
c.) Temperature dependence of the cavity frequency. It decreases towards higher temperatures. The total decrease is less than $0.4\,\%$ and thus very small compared to the temperature dependence of the coupling strength $g(T)$. The error bars are of the order of $0.1\,\mathrm{mT}$. 
}
\label{res_1}
\end{figure}

In order to explain the temperature dependence of the coupling strength by the change in the spin number $N$ and thus the saturation magnetization, the contribution of the other quantities in the expression for $g_{\rm{eff}}(T)$ (Eq. \ref{Eq:coupling_strength})  has to be considered. In Fig. \ref{res_1}, the temperature dependent changes of the applied external field $H_{\rm{ext}}$ at resonance (\ref{res_1} a.)), the gyromagnetic ratio $\gamma$  (\ref{res_1} b.)) and the resonance frequency of the cavity resonator $\omega_{\rm{r}}$ (\ref{res_1} c.)) are displayed. 
The  external field $H_{\rm{ext}}$ changes by $4\,\%$, the value for the  gyromagnetic ratio $\gamma$ by $4\,\%$ following the changes in the field values and  resonance frequency of the cavity resonator $\omega_{\rm{r}}$ by $0.5\, \%$, respectively.  These changes are alltogether much smaller than the absolute change in the coupling strength from room temperature to the low temperature regime. Therefore, the change in the coupling strength is governed by the variation of the participating spin number $N$ for the particular magnetostatic mode.
\subsection*{Field dependence of the magnon linewidth $\kappa_{\rm{m}}/2\pi(T)$ close to resonance  }
\begin{figure}[h!]
\centering
\includegraphics[scale=0.41]{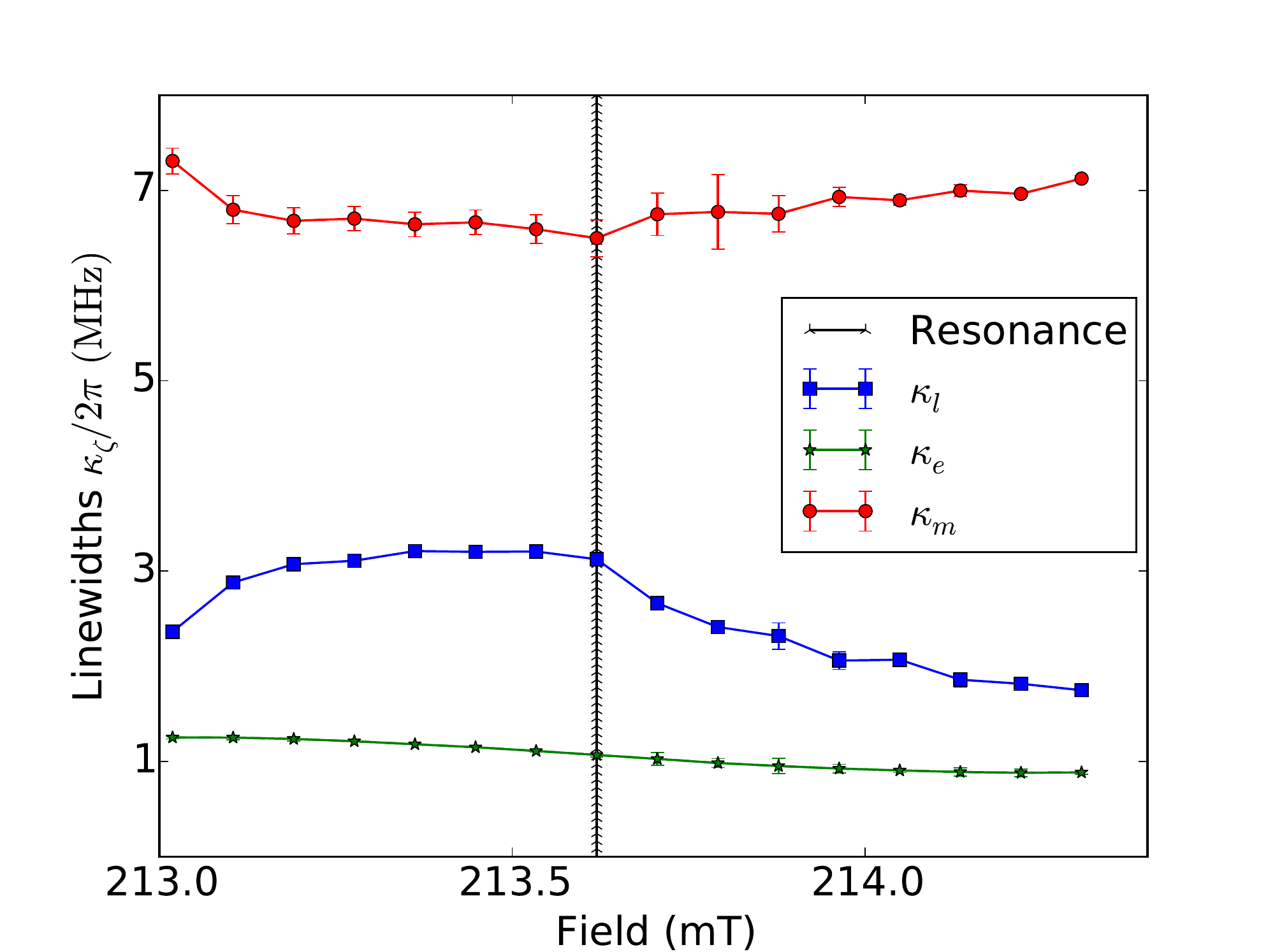}
\caption{Typical field dependence of the linewidth parameters at $T= 50\,\mathrm{K}$. The datapoints are taken from a reflection measurement. The red curve (circles) shows values for the magnon linewidth $\kappa_{\rm{m}}$, the blue curve (squares)  the linewidth related to the resonator, labeled with $\kappa_{\rm{r}}$, and the green one (stars), labeled with 
$\kappa_{\rm{e}}$, the field dependence of the coupling of the resonator to the microwave feedline. Since the magnon linewidth is dominated by scattering within the sphere, it does not depend on the resonance condition. Indeed, both $\kappa_{\rm{m}}$ and 
$\kappa_{\rm{e}}$ are constant within the error bars for the displayed range of applied external field. Approaching resonance opens a new loss channel due to the energy transfer to the magnons in the YIG sphere. The total cavity losses $\kappa_{\rm{l}}$ increase by 
$50\,\%$ when reaching the resonance condition.}
\label{Fig6_kappa_resonance}
\end{figure}

When the Kittel magnons are on resonance with the cavity, energy is transferred to the Kittel magnons. Thus, approaching the resonance condition for the magnon polariton, $\omega_{\rm{r}} \equiv \omega_{\rm{m}}$, adds loss to the cavity. Accordingly, the total losses of the cavity resonator increases to a maximum value at resonance. As discussed in the main part, the magnonic losses are due to internal scattering processes. Likewise the external losses from the feedline coupling do not depend on the resonance condition, as shown in Fig. \ref{Fig6_kappa_resonance}.
\subsection*{Evolution of difference between fields of main and supressed avoided crossings}
\begin{figure}[h!]
\centering
\includegraphics[scale=0.4]{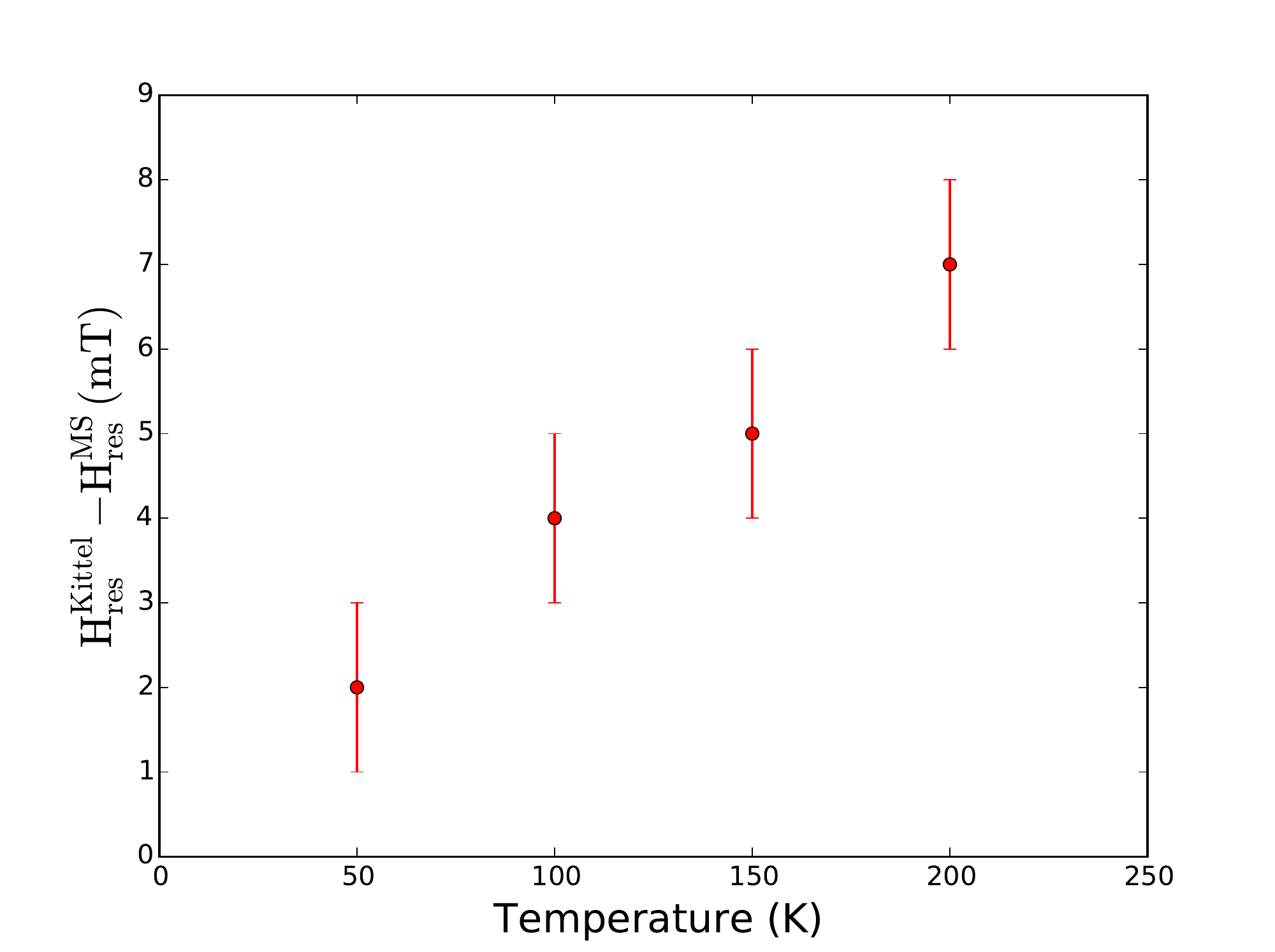}
\caption{Difference between the applied field for resonant coupling of the cavity photon to the Kittel mode ($\rm{H_{res}^{Kittel}}$) and to other magnetostatic modes ($\rm{H_{res.}^{MS}}$) in the specimen, respectively. As shown exemplary in the left inset of Fig. \ref{gMs}, the difference is estimated from the distance of the resonant field for coupling to the Kittel mode to the field for the additional coupling to other magnetostatic modes.  
For lower temperatures, the probability for parasitic coupling to other modes increases and the coupling strength of the Kittel mode is lowered (c. f. Fig. \ref{gMs}). Above $200\,\mathrm{K}$, the distance is large and thus the parasitic coupling rate decreases below the noise level. For temperatures below $50\,\mathrm{K}$, the resonance frequencies are almost identical. At the same time, the overall signal-to-noise ratio is reduced and the field difference cannot be resolved.}
\label{Fig.7_distance}
\end{figure}

The data shown in Fig. \ref{Fig.7_distance} reveals how the difference in applied external field between the Kittel mode anticrossing and the most prominent, supressed one, decreases for  lower  temperatures. As mentioned in the main text, this leads to additional coupling to other magnetostatic modes in the YIG sphere and decreases the coupling strength value for a coupling to the Kittel mode.
\subsection*{Temperature evolution of the relative quantity $g_{\rm{eff}}(T)/\sqrt{M_{\rm{s}}}$}
The following figure, Fig. \ref{Fig_8_ratio}, shows, that the ratio between the  temperature dependent change in the coupling strength and the square root of the saturation magnetization, as $M_{\rm{s}}\propto N$ does not change significantly within the error bars.  If another quantity would have a comparable influence on the change in the coupling strength, the value of this relative quantity would change more. 
Thus, together with comparing the absolute changes of all quantities (Fig. \ref{gMs} and Fig.\ref{res_1} a.)-c.)) contributing to the coupling strength (Eq. \ref{Eq:coupling_strength}), the change in the spin number $N$ as a function of temperature can be considered as the dominant one, in general. However, for temperatures lower than $100\,\mathrm{K}$, additional coupling besides the Kittel mode, decreases the coupling strength. 
\begin{figure}[h!]
\centering
\includegraphics[scale=0.4]{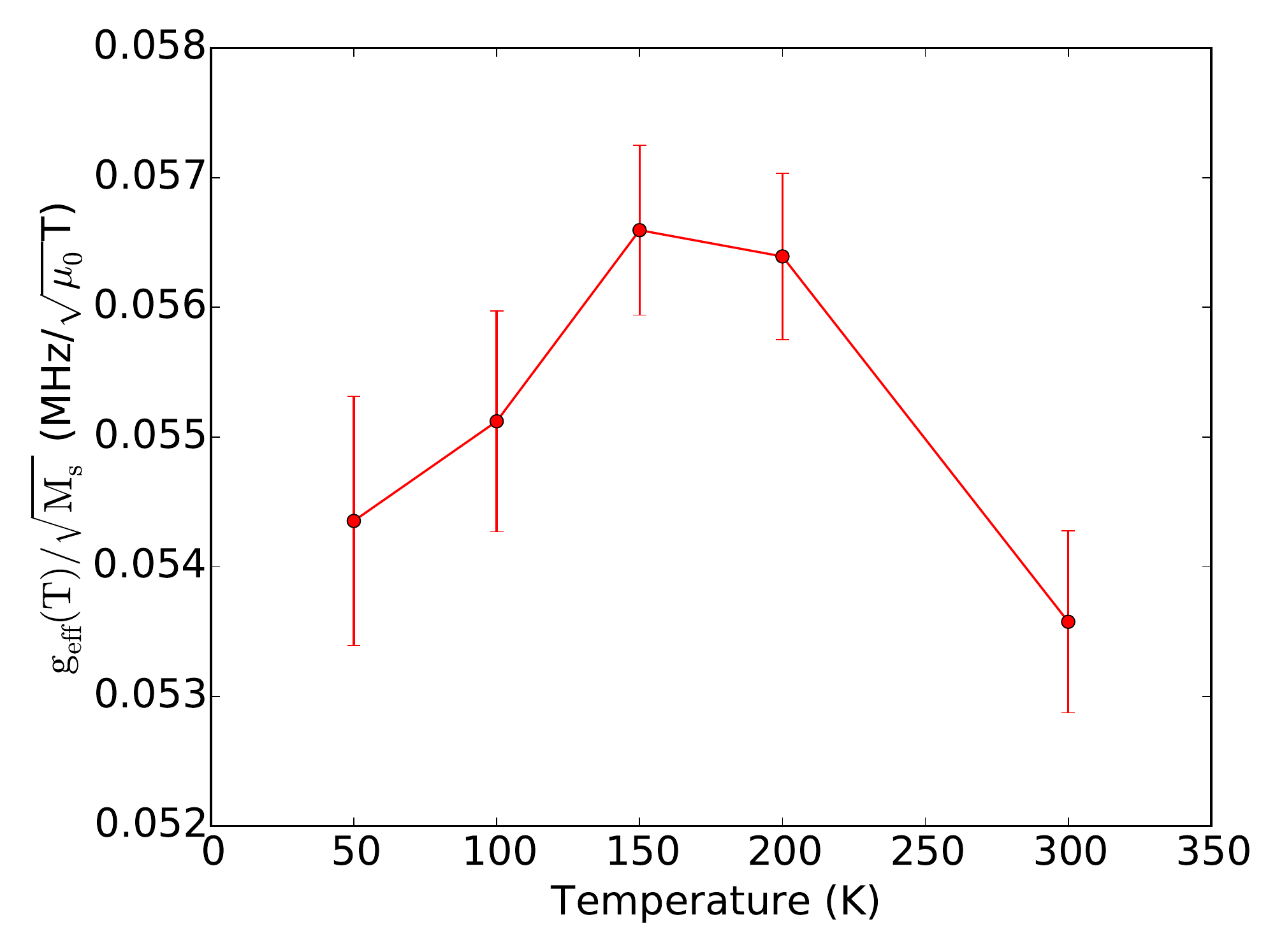}
\caption{Ratio between the coupling strength $g_{\rm{eff}}(T)$ and the square root of the magnetization $M_{\rm{s}}$. Over the whole temperature range the value of that relative quantity changes by $6\,\%$.}
\label{Fig_8_ratio}
\end{figure}
\end{document}